\documentstyle[eqsecnum,aps,preprint,epsfig]{revtex}
\begin{document}
\draft
\title{Spin fluctuations in nearly magnetic metals
from ab-initio dynamical spin susceptibility calculations:
application to $Pd$ and $Cr_{95}V_{5}$}
\author{J.B.Staunton\dag, J.Poulter\ddag, B.Ginatempo\P, E.Bruno\P and
D.D.Johnson\S}
\address{\dag\ Department of Physics, University of Warwick, Coventry CV4~
7AL, U.K.}
\address{\ddag\ Department of Mathematics,Faculty of Science,
Mahidol University,Bangkok 10400,
Thailand.}
\address{\P\ Dipartimento di Fisica and Unita INFM, Universita di Messina, Italy}
\address{\S\ Department of Materials Science and Engineering, University of
Illinois, IL 61801, U.S.A.}
\date{\today}
\maketitle
\begin{abstract}
We describe our theoretical formalism and computational
scheme for making {\it ab-initio} calculations of the dynamic
paramagnetic spin
susceptibilities of metals and alloys at finite temperatures.
Its basis is Time-Dependent Density Functional Theory within an
electronic multiple
scattering, imaginary time Green function formalism. Results
receive a natural interpretation in terms of overdamped
oscillator systems making them suitable for
incorporation into spin fluctuation theories. For illustration we
apply our method to the nearly ferromagnetic
metal $Pd$ and the nearly antiferromagnetic chromium  alloy
$Cr_{95}V_{5}$. We compare and contrast the spin dynamics of
these two metals and in each case identify those
fluctuations with relaxation times much longer than
typical electronic `hopping times'.
\end{abstract}
There is renewed interest in spin fluctuations in materials
close to magnetic order. This is due in part to a realisation
  that nearly critical magnetic fluctuations may be important factors
governing the
  non-conventional properties of a wide range of
  materials which include the
  high $T_{c}$ superconducting cuprates and heavy fermion systems
  \cite{Sachdev}.
  The strongly correlated electrons in many of these systems, however,
  have meant that most theoretical work has concentrated on
parameterised models
  in which the electronic motion is treated rather simply. Another
complementary approach is
  to use an ab initio theory such as Time Dependent Density Functional
  Theory  (TDDFT) \cite{Gross_review}
  but apply it to materials where it can be expected
  to work i.e. where the effects of electron
  correlations are not so important, but which otherwise have
  important similarities to
  the systems in question. For example, with its perovskite
structure containing
  transition metal(TM)-oxygen
  planes, $Sr_{2}RuO_{4}$ has several aspects in common with
  the HTC materials. But the presence of the 4d TM $Ru$ rather than
the
  narrower band 3d TM $Cu$ means that electron correlation effects are
smaller and
  therefore DFT-based calculations can provide a valuable starting
point. Moreover, its
exotic superconductivity at low temperature
  seems likely to be affected by spin fluctuations
\cite{pwave,Singh,MP}.
  Concerning another example, the transition temperature separating
paramagnetic and
  magnetically ordered phases of the
  cubic transition metal compound $MnSi$, which has the B20 crystal
structure,
  is driven down to zero temperature
  upon the application of pressure. In the vicinity
  of the critical pressure for this quantum phase transition the
system exhibits
  non- or marginal Fermi liquid properties \cite{GGLetc}. In this
paper we investigate the energy and
temperature dependence of spin fluctuations in
two systems which although structurally and compositionally simpler
than the two mentioned above, are
also close to magnetic phase transitions. The first is palladium
which is nearly ferromagnetic and the second is the nearly
anti-ferromagnetic  $Cr_{95}V_{5}$ alloy.

Theoretical models
  in which an effective action for the slow spin fluctuations is
written down
  have contributed greatly to our understanding of the properties
of itinerant
  electron systems close to magnetic order \cite{GGLetc,MD+Moriya+LT}.
  Recent work which can incorporate results from DFT-based `Fixed Spin
Moment'
  (FSM) electronic structure calculations, treats these
fluctuations classically
  \cite{FSM-SF}. A Landau-Ginzburg-like energy
  functional is written down, a free
  energy constructed which includes terms describing the
interactions between
  the fluctuations (the mode-mode coupling) and properties such
  as the static susceptibility, specific heat and resistivity
calculated.
  The FSM electronic
  structure calculations can be used to determine the
  coefficients in this functional \cite{FSM-SF}. Many informative
  studies have been carried out.
  For those calculations with such a DFT basis, these still
  remain qualitative investigations because of the lack of a
prescription for the
  effective
  number of modes to include in the theory and its variation with
temperature
  \cite{GGLetc,MD+Moriya+LT}. Stoner single-particle excitation
effects are also largely ignored. Both these issues can be addressed
by the development and application of methods to calculate the
temperature-dependent, dynamic paramagnetic spin susceptibility of
nearly magnetic materials. The development of dynamic susceptibility
calculations is particularly
  pertinent now that  inelastic neutron scattering experiments,
  such as the time of flight measurements,
 have developed  to the extent that spin fluctuations in
nearly
  magnetic metals can be accurately measured \cite{Hayden}.

We have recently devised and proven a new scheme for calculating
  the wave-vector and frequency dependent dynamic spin susceptibility
of
  metallic systems \cite{dyn} which
  is based on the Time Dependent Density Functional Theory (TDDFT) of
Gross et al.
  \cite{Gross_review} and as such is an all electron theory. This
enables the
    temperature dependent dynamic spin susceptibility of metals and
  compositionally disordered alloys to be calculated.

Over the past few years great progress has been made
in establishing TDDFT \cite{Gross_review}. Analogs of the
Hohenberg-Kohn \cite{HK+KS} theorems of the static density
functional (DFT) formalism
have been proved and rigorous properties found.
By considering a
paramagnetic metal subjected to a small, time-dependent
external magnetic field, ${\bf b}({\bf r},t)$ which induces a
magnetisation ${\bf m} ({\bf r},t)$ and using
TDDFT in \cite{dyn} an expression for the dynamic
paramagnetic spin susceptibility $\chi ({\bf q}, w)$ via
a variational linear response approach can be derived \cite{Yang}.
Accurate calculations of dynamic susceptibilities from this
basis have been scarce (e.g. \cite{SW+Sav}) because
they are difficult and computationally demanding. In ref.\cite{dyn}
we showed that these
problems can be mitigated by  accessing $\chi ({\bf q}, w)$ via the
corresponding {\it temperature} susceptibility $\bar{\chi} ({\bf q}, w_{n})$
where $w_{n}$ denotes a bosonic Matsubara frequency \cite{Fetter+Walecka}. We demonstrated this approach by an
investigation into the nature of the
  spin fluctuations in paramagnetic $Cr$ and compositionally
disordered
    $Cr_{95}V_{5}$ and $Cr_{95}Re_{5}$ alloys with good agreement with
  experimental data \cite{dyn}. For example, recent inelastic neutron
scattering
    experiments \cite{Hayden,Fawcett} have measured incommensurate AF
    `paramagnons', persisting up to high frequencies in the latter
system which
  were also shown in our calculations

 In \cite{dyn} we provided a brief outline of
this approach for crystalline systems with one
atom per unit cell only
so in the next section we provide its
generalisation to multi-atom per unit cell materials in which one
or more sublattices may be substitutionally disordered. We also
describe the techniques involved in the method in more detail. In the
following section we discuss our study of the frequency,
wave-vector and temperature dependence of the spin fluctuations in
$Pd$ and identify the paramagnon regime. We then move on to carry out
a similar study of the incommensurate antiferromagnetic spin
fluctuations in the dilute $Cr_{95}V_{5}$ alloy.
In each case the results can be interpreted simply in terms of
an over-damped harmonic oscillator model \cite{GGLetc}. In the
latter case we show how the parameters scale with temperature
and compare and contrast the two systems.
In the final section we draw some conclusions
and remark how this work has the potential to be integrated into spin
fluctuational theories of nearly and weakly magnetic metals
\cite{GGLetc}.

\section{The dynamical spin susceptibility $\chi ({\bf q}, w)$.}

The equilibrium state of a paramagnetic metal, described by standard DFT,
has density $\rho_{0}({\bf r})$ and its
magnetic response function
\begin{equation}
\chi ({\bf r} t;{\bf r}^{\prime}t^{\prime}) = \frac
{\delta m [b] ({\bf r},t)}{\delta b ({\bf
r}^{\prime},t^{\prime})}|_{b=0,\rho_{0}}
\end{equation}
 is given by the following Dyson-type equation.
\begin{equation}
\chi ({\bf r}t;{\bf r}^{\prime}t^{\prime}) = \chi_{s} ({\bf r} t;{\bf
r}^{\prime} t^{\prime}) + \int d{\bf r}_{1} \int d t_{1} \int d{\bf r}_{2} \int
d t_{2} \chi_{s} ({\bf r} t;{\bf r}_{1}t_{1}) K_{xc}({\bf r}_{1} t_{1}; {\bf
r}_{2} t_{2}) \chi ({\bf r}_{2} t_{2}, {\bf r}^{\prime} t^{\prime}) \nonumber
\end{equation}
$\chi_{s}$ is the magnetic response function of the Kohn-Sham
non-interacting system with the same unperturbed density $\rho_{0}$ as the full interacting electron system, and
\begin{equation}
K_{xc}({\bf r} t;{\bf r}^{\prime}t^{\prime})
= \frac{\delta b_{xc} ({\bf r},t)}{\delta m ({\bf
r}^{\prime},t^{\prime})}|_{b=0,\rho_{0}}
\end{equation}
is the functional derivative of the effective
exchange-correlation magnetic field with respect to the induced magnetisation.
As emphasised in ref.\cite{Gross_review} eq.1.2 represents an exact
representation  of the linear magnetic response. The corresponding development for
systems at finite temperature in thermal equilibrium has also been described
\cite{Yang}. In practice
approximations to $K_{xc}$ must be made and this work employs the adiabatic
local approximation (ALDA) \cite{Gross_review} so
that
\begin{eqnarray}
K^{ALDA}_{xc} ({\bf r} t;{\bf
r}^{\prime} t^{\prime}) &
= & \frac{d b_{xc}^{LDA}(\rho({\bf r},t),m({\bf r}, t))}{
dm({\bf r},t)}|_{\rho_{0},m=0} \delta ({\bf r} -{\bf
r}^{\prime}) \delta
(t-t^{\prime}) \nonumber \\ & = &
I_{xc}({\bf r}) \delta ({\bf r} -{\bf r}^{\prime}) \delta
(t-t^{\prime})
\end{eqnarray}
 On taking the Fourier transform of eq.1.2 with
respect to time we obtain the
dynamic spin susceptibility $\chi ({\bf r}, {\bf r}^{\prime}; w)$.

For computational expediency we consider the corresponding {\it temperature}
susceptibility \cite{Fetter+Walecka} $\bar{\chi} ({\bf r}, {\bf r}^{\prime};
w_{n})$ which occurs in the Fourier representation of the temperature function
$\bar{\chi} ({\bf r} \tau;{\bf r}^{\prime} \tau^{\prime})$ that depends on
imaginary time variables $\tau$,$\tau^{\prime}$ and $w_{n}$ are the bosonic
Matsubara frequencies $w_{n}=
2 n \pi k_{B} T$. Now $\bar{\chi}( {\bf r}, {\bf
r}^{\prime}; w_{n}) \equiv \chi ({\bf r}, {\bf r}^{\prime}; i w_{n})$
and an
analytical continuation to the upper side of the real $w$ axis produces the
dynamic susceptibility $\chi ({\bf r}, {\bf r}^{\prime}; w)$.

We define our general system as having a crystal structure with
lattice vectors $\{ {\bf R}_{i} \}$ and where there are
$N_{s}$ non-equivalent atoms per unit cell. On the $l^{th}$ of
the $N_{s}$ sublattices there are $N_{l}$ atomic species with
concentrations $c^{(l)}_{\alpha_{l}}$
($\alpha_{l}= 1,\cdots,N_{l}$). In
each unit cell the $N_{s}$ atoms are situated at locations ${\bf
a}_{l}, l=1,\cdots,N_{s}$. The volume of the unit cell is $V_{WS}$.
On carrying out a lattice Fourier transform over the lattice vectors
$\{ {\bf R}_{i} \}$ we obtain the
following Dyson equation for the temperature susceptibility
\begin{eqnarray}
\bar{\chi}^{\alpha_{l}}({\bf
x}_{l},{\bf x}^{\prime}_{l^{\prime}},{\bf q},w_{n}) & = &
\sum_{\gamma_{l^{\prime}}}^{N_{l^{\prime}}}
c^{l^{\prime}}_{\gamma_{l^{\prime}}}
\bar{\chi}_{s}^{\alpha_{l} \gamma_{l^{\prime}}}({\bf
x}_{l},{\bf x}_{l^{\prime}}^{\prime},{\bf
q},w_{n}) \nonumber \\ & & + \sum_{l^{\prime \prime}}^{N_{s}} \int
d {\bf x}^{\prime \prime}_{l^{\prime \prime}}
\sum_{\gamma_{l^{\prime \prime}}}^{N_{l^{\prime \prime}}}
c^{l^{\prime \prime}}_{\gamma_{l^{\prime \prime}}}
\bar{\chi}_{s}^{\alpha_{l} \gamma_{l^{\prime \prime}}}({\bf
x}_{l},{\bf
x}^{\prime \prime}_{l^{\prime \prime}},{\bf
q},w_{n}) I_{xc}^{\gamma_{l^{\prime \prime}}} ({\bf
x}^{\prime \prime}_{l^{\prime
\prime}}) \bar{\chi}^{\gamma_{l^{\prime \prime}}}({\bf
x}^{\prime \prime}_{l^{\prime \prime}}, {\bf
x}^{\prime}_{l^{\prime}},{\bf q},w_{n})              \label{eq:temp}
\end{eqnarray}
with ${\bf x}_{l}$,${\bf x}^{\prime}_{l^{\prime}}$
and ${\bf x}^{\prime \prime}_{l^{\prime \prime}}$ measured relative to
atomic cells centred on ${\bf a}_{l}$, ${\bf a}_{l^{\prime}}$ and
${\bf a}_{l^{\prime \prime}}$ respectively. $\bar{\chi}_{s}^{\alpha_{l}
\gamma_{l^{\prime}}}$ describes the non-interacting susceptibility
{\it averaged
over all configurations} in which one site on sublattice $l$ is occupied
by an $\alpha_{l}$ atom and another site on sublattice $l^{\prime}$ is
occupied by a $\gamma_{l^{\prime}}$ atom. $\bar{\chi}^{\alpha_{l}}$ is the
full susceptibility averaged over all configurations where an $\alpha_{l}$
atom is located on one site on sublattice $l$.

In terms of the lattice Fourier transform of the DFT Kohn-Sham
Green function of the static unperturbed system the non-interacting
susceptibility can be written approximately
\cite{Butler}
\begin{eqnarray}
 \bar{\chi}_{s}^{\alpha_{l}
\gamma_{l^{\prime}}}({\bf x}_{l},{\bf x}^{\prime}_{l^{\prime}},{\bf
q},w_{n})& = & \nonumber \\
  -\frac{1}{\beta} \sum_{m} \int \frac{
d{\bf k}}{v_{BZ}} & &\langle G({\bf
x}_{l},{\bf x}^{\prime}_{l^{\prime}}, {\bf k}, \mu
+i\nu_{m}) \rangle_{\alpha_{l}\gamma_{l^{\prime}}} \langle G({\bf
x}_{l^{\prime}}^{\prime},{\bf x}_{l},{\bf k}+{\bf q},\mu + i(
\nu_{m}+ w_{n}))\rangle_{\gamma_{l^{\prime}} \alpha_{l}}  \label{eq:G}
\end{eqnarray}
where the integral is over the Brillouin zone of the lattice and ${\bf
k}$, ${\bf q}$ and ${\bf k}+{\bf q}$ are all wavevectors within this
Brillouin zone which has volume $v_{BZ}$. $\mu$ is the chemical
potential and $\nu_{m}$ is a fermionic Matsubara
frequency $(2n+1) \pi k_{B} T$.

The Green function can be obtained within the
framework of multiple scattering (KKR) theory
\cite{Faulkner+Stocks} and this makes this formalism applicable to
disordered alloys as well as ordered compounds and elemental metals, the
disorder being treated by the Coherent Potential Approximation (CPA)
\cite{KKRCPA}. Then the partially averaged Green function with species
$\alpha_{l}$ on a site at ${\bf R}_{i} + {\bf a}_{l}$
 on the $l^{th}$ sublattice and species
$\gamma_{l^{\prime}}$ on a site at ${\bf R}_{j} + {\bf
a}_{l^{\prime}}$ on sublattice number $l^{\prime}$
\begin{equation}
\langle G ({\bf R}_{i}+{\bf a}_{l} +{\bf x}_{l} ,{\bf
R}_{j} + {\bf a}_{l^{\prime}} + {\bf x}_{l^{\prime}},z)
\rangle_{ \alpha_{l}, \gamma_{l^{\prime}}}
\end{equation}
can be evaluated in terms
of deviations from the Green function of an electron propagating through a lattice
of identical potentials determined by the CPA ansatz \cite{Butler}.
The lattice Fourier transform used in equation ~(\ref{eq:G}) is
expressed as
\begin{eqnarray}
\langle G({\bf x}_{l},{\bf x}^{\prime}_{l^{\prime}}, {\bf k}, z)
\rangle_{\alpha_{l}\gamma_{l^{\prime}}} & = & \sum_{L,L^{\prime}}
Z_{L}^{\alpha_{l}} ({\bf x}_{l},z) \langle
\tau^{l,l^{\prime}}_{L,L^{\prime}} ({\bf k},z)
\rangle_{\alpha_{l}\gamma_{l^{\prime}}}
Z^{\gamma_{l^{\prime}}}_{L^{\prime}}({\bf x}^{\prime}_{l^{\prime}},z)
\nonumber \\& + &
\delta_{l,l^{\prime}} \sum_{L
,L^{\prime}} [ \delta_{\alpha_{l}\gamma_{l}} (  Z_{L}^{\alpha_{l}}
({\bf x}_{l},z)  \langle
\tau^{l,l}_{L,L^{\prime}} (z) \rangle_{\alpha_{l}}
Z^{\alpha_{l}}_{L^{\prime}}({\bf x}^{\prime}_{l},z)
- Z_{L}^{\alpha_{l}} ({\bf x}_{l}^{<},z) J_{L}^{\alpha_{l}} ({\bf
x}_{l}^{>},z)) \nonumber \\& - & Z_{L}^{\alpha_{l}} ({\bf x}_{l},z)
\langle \tau^{l,l}_{L,L^{\prime}} (z) \rangle_{\alpha_{l}
\gamma_{l}} Z^{\gamma_{l}}_{L^{\prime}}({\bf x}^{\prime}_{l},z)]
\end{eqnarray}
where $Z_{L}^{\alpha_{l}}$ and $J_{L}^{\alpha_{l}}$ are
respectively the regular and irregular solutions of the Schrodinger
equation in an atomic cell on sublattice $l$ inhabited by an
$\alpha_{l}$ atom and $L$,$L^{\prime}$ represent the angular momentum
quantum numbers. The lattice Fourier transform of the
averaged scattering
path operator $ \tau^{l,l^{\prime}}_{L,L^{\prime}} ( {\bf R}_{i} -{\bf
R}_{j}, z)$ is specified by the
following expressions \cite{Faulkner+Stocks}
\begin{equation}
\langle \tau^{l,l^{\prime}}_{L,L^{\prime}} ({\bf k},z)
\rangle_{\alpha_{l}\gamma_{l^{\prime}}} = \sum_{L_{1} L_{2}}
D^{l}_{\alpha_{l},L L_{1}} (z) \bar{\tau}^{l,l^{\prime}}_{L_{1},L_{2}}
({\bf k},z) \tilde{D}^{l^{\prime}}_{\gamma_{l^{\prime}}, L_{2}
L^{\prime}}
\end{equation}
where, in matrix notation with respect to angular momentum
$L$,($\tilde{D}$ describes the transpose)
\begin{equation}
D^{l}_{\alpha_{l},L L_{1}} (z) = [ 1 + \bar{\tau}^{l l}(z)
(m^{\alpha_{l}}(z) - m^{c,l}(z))]^{-1}_{L L_{1}}
\end{equation}
with $m^{\alpha_{l}}$ being the inverse of the scattering t-matrix
$t^{\alpha_{l}}$ of an $\alpha_{l}$ atom potential, $m^{c,l}$ the
inverse of the CPA t-matrix for the $l^{th}$ sublattice, $t^{c,l}$ and
$\bar{\tau}^{l l^{\prime}}$ the unit cell-diagonal part of
the CPA scattering path operator. The following averages also involve
these quantities
\begin{eqnarray}
\langle \tau^{l,l^{\prime}}_{L,L^{\prime}}
(z)\rangle_{\alpha_{l}\gamma_{l^{\prime}}} & = &
\sum_{L_{1} L_{2}}   D^{l}_{\alpha_{l},L L_{1}} (z)
\bar{\tau}^{l,l^{\prime}}_{L_{1},L_{2}}(z)\tilde{D}^{l^{\prime}}_
{\gamma_{l^{\prime}}, L_{2} L^{\prime}} (z) \nonumber \\
\langle \tau^{l,l}_{L,L^{\prime}} (z)\rangle_{\alpha_{l}} & = &
\sum_{L_{1}} D^{l}_{\alpha_{l},L L_{1}}
(z) \bar{\tau}^{l,l}_{L_{1},L^{\prime}} (z)
\end{eqnarray}
Finally the lattice Fourier transform of
this CPA scattering path operator is given by
\begin{equation}
\bar{\tau}^{l l^{\prime}}_{L,L^{\prime}} ({\bf k},z) =
t^{c,l}_{L,L^{\prime}} (z) \delta_{l l^{\prime}} + \sum_{l^{\prime
\prime}} \sum_{L_{1} L_{2}} t^{c,l}_{L,L_{1}}(z) g^{l l^{\prime
\prime}}_{L_{1} L_{2}} ({\bf k},z) \bar{\tau}^{l^{\prime \prime}
l^{\prime}}_{L_{2} L^{\prime}} ( {\bf k},z)
\end{equation}
and $g^{l l^{\prime}}_{L_{1} L_{2}} ({\bf k},z)$ are the structure
constants for the crystal structure \cite{KKRbook}.

To solve equation ~(\ref{eq:temp}), we use a direct method of
matrix inversion in which, for example, $\bar{\chi}_{s}$ is cast into matrix form of order $(\sum_{l=1}^{N_{s}} S_{l} N_{l})\times
(\sum_{l=1}^{N_{s}} S_{l} N_{l})$ where
$S_{l}$ is the number of spatial grid points associated with the
$l^{th}$ sublattice. Local field effects are thus fully incorporated.
The full Fourier transform
\begin{equation}
\bar{\chi}({\bf q},{\bf q}; w_{n})=(1/V_{WS}) \sum_{l}
\sum_{\alpha_{l}} c^{l}_{\alpha_{l}}
\sum_{l^{\prime}} e^{i{\bf q} \cdot({\bf a}_{l} - {\bf
a}_{l^{\prime}})} \int d {\bf x}_{l} \int
d {\bf x}^{\prime}_{l^{\prime}} e^{i{\bf q} \cdot({\bf
x}_{l} - {\bf x}^{\prime}_{l^{\prime}})} \bar{\chi}^{\alpha_{l}}({\bf
x}_{l},{\bf x}^{\prime}_{l^{\prime}}, {\bf q}, w_{n})
\end{equation}
 can then be constructed.

The most computationally demanding parts of the calculation are the convolution
integrals over the Brillouin Zone which result from the
expression  for $\bar{\chi}_{s}$, eq.~(\ref{eq:G}) i.e.
\begin{equation}
\int d{\bf k} \bar{\tau}^{l l^{\prime}}_{L L_{1}} ({\bf k},z)
\bar{\tau}^{l^{\prime} l}_{L_{2} L_{3}}({\bf k}+{\bf q},z^{\prime})
\end{equation}
Since all electronic structure quantities are
evaluated at complex energies $z$, these convolution integrals have no
sharp structure and can be evaluated
straightforwardly by an application of the adaptive grid method
of B.Ginatempo and E.Bruno
\cite{Ben+Ezio} which has been found to be very
efficient and accurate. In this method one can preset the level of
accuracy of the integration by supplying an error parameter
$\epsilon$. In the calculations described in this paper we have
used $\epsilon = 10^{-4}$ so that we obtain 4
significant figure accuracy on $\chi_{s}$. This control on accuracy
is crucial for the description of the long wavelength paramagnons
in $Pd$. In this case we needed to sample around 2000 ${\bf k}$-points
at energies on the contour distant 0.5Ry. off the real axis but up to
50000 ${\bf k}$-points very close (0.001 Ry.) to the real
energy axis. Around half that number of points were required for
the $Cr_{95}V_{5}$ calculations.

We evaluate
the Matsubara sum shown in equation ~(\ref{eq:G}) by
using suitable complex energy contours to enclose some of the
Matsubara poles and exploiting the hermiticity of the single-electron
Green function $G ({\bf r},{\bf r}^{\prime},z)$. Consequently a
quantity like $\chi_{s}$ at a positive bosonic Matsubara frequency
$w_{n}$ with form
\begin{equation}
A({\bf r},{\bf r}^{\prime}, w_{n}) = -\frac{1}{\beta} \sum_{m}
G({\bf r},{\bf r}^{\prime},\mu + i \nu_{m}) G({\bf r}^{\prime},{\bf
r}, \mu + i \nu_{m} + i w_{n})
\end{equation}
where the fermionic
Matsubara sum is over $m= 0,\pm 1,\pm 2, \cdots, \pm \infty$ can be
rewritten as
\begin{eqnarray}
A({\bf r},{\bf r}^{\prime}, w_{n})& = & -\frac{2}{\beta}
\sum_{m=0}^{M_{max}} Re( G({\bf r},{\bf r}^{\prime},\mu
+ i \nu_{m}) G({\bf r}^{\prime},{\bf
  r}, \mu + i \nu_{m} + i w_{n}) ) + \\
\frac{1}{\pi} Im \int_{C}
dz & (G &({\bf
r},{\bf r}^{\prime}, z) G({\bf r}^{\prime},{\bf r},z+ i w_{n} )) f(z)
-\frac{1}{\beta} \sum_{m=0}^{n-1}
G({\bf r},{\bf r}^{\prime}, \mu -i \nu_{m}) G({\bf
r}^{\prime},{\bf r}, \mu + i w_{n} -i \nu_{m}) \nonumber
\end{eqnarray}
Both sums are over positive fermionic frequencies only and $f(z)$
is the Fermi function. $C$ is a
contour which encloses the first $M_{max}$ fermionic Matsubara
poles with the real axis. In our calculations we have used a simple
box contour whose top lies exactly halfway between
 two neighbouring $\nu_{m}$'s which makes the Fermi factor at this $z$ real.
The
terms in the Matsubara sums for $\nu_{m}$'s $> 0.001$ are evaluated by
interpolating from a grid $\nu = 0.001\times 10^{0.1 p}$, with
$p=0,1,\cdots,30$ or are evaluated directly. $M_{max}$ is chosen so
$\nu_{M_{max}} \approx 1$.

Once the temperature susceptibility $\bar{\chi}({\bf q},{\bf q};
w_{n})$ has been calculated the dynamic susceptibility can be found.
As discussed in ref. \cite{Fetter+Walecka}, for example, we can define the retarded
response function $\chi ({\bf q},{\bf q}, z)$ of a complex variable $z$. Since
it can be shown  \cite{Fetter+Walecka}
formally that $\lim_{z \rightarrow \infty} \chi (z) \sim 1/z^{2}$ and we can
obtain $\chi (i w_{n})$ from the above analysis it is possible to continue
analytically to values of $z$ just above the real axis, i.e. $z= w +i \eta$.
In order to achieve this we fit our data to a function describing a set of
overdamped oscillators, i.e.,
\begin{equation}
\bar{\chi} ({\bf q}, {\bf q},w_{n}) = \sum_{M}
\frac{ \chi_{M}({\bf q})} {(1+(w_{n}/\Gamma_{M}({\bf
q}))+(w_{n}/\Omega_{M}({\bf q}))^{2})} \label{eq:fit}
\end{equation}
in which $M$ is an integer. From this the dynamic susceptibility
can be written down and the imaginary part becomes
\begin{equation}
Im \chi({\bf q},{\bf q},w) = \sum_{M} \frac{ \chi_{M}({\bf q})
(w/\Gamma_{M}({\bf q}))}{\{[1-(w/\Omega_{M}({\bf q}))^{2}]^{2}
+(w/\Gamma_{M}({\bf q}))^{2}\}}.
\end{equation}
This form ensures that the sum rule
involving the static susceptibility $\chi ({\bf q})$ is satisfied, i.e.
\begin{equation}
\chi
({\bf q})= \frac{2}{\pi}
\int^{\infty}_{0} dw  \frac {Im \chi({\bf q},{\bf q},w)}{w}
\end{equation}
In nearly magnetic metals the nature of the spin fluctuations
can be succinctly described via the variance $<m^{2}({\bf q})>$.
From the fluctuation dissipation theorem
\begin{equation}
<m^{2}({\bf q})> = \frac{1}{\pi} \int^{\infty}_{-\infty} dw
\frac{Im \chi({\bf q},{\bf q},w)}{ (1-e^{-\beta w})}
\end{equation}
By introducing an energy cutoff in this expression the slower
fluctuations can be revealed.

We find that very good fits are obtained with small $M$ in
equation ~(\ref{eq:fit}) for a wide
range of $w_{n}$'s ( $ < 500$ meV) for the systems we have examined
to date. For $Cr$ and its dilute alloys using $M=1$ in
equation ~(\ref{eq:fit}) provides an excellent fit. This means
that a close
analogy with an overdamped harmonic oscillator can be made. In the
case of palladium, equation ~(\ref{eq:fit}) with $M=2$ gives an
excellent representation of $\bar{\chi} ({\bf q}, {\bf q},w_{n})$.
Now the spin fluctuations can be described in terms of two
overdamped oscillators with two distinct timescales. The slower
one describes the famous paramagnons of palladium.

\section{The spin fluctuations of nearly ferromagnetic palladium.}

Although possessing one of the highest Fermi energy densities
of states of all the transition metals, palladium has neither a
magnetically ordered nor a superconducting ground state. With
its nearly filled d-bands, it is,
however, on the verge of being a ferromagnet. This is shown by its
uniform paramagnetic spin susceptibility
which is greatly enhanced over that of an equivalent
non-interacting system by exchange-correlation effects.
Moreover, the temperature dependence of the susceptibility also
demonstrates effects of long-lived ferromagnetic
spin-fluctuations or paramagnons. It has even been
proposed that these spin fluctuations might induce p-wave
superconductivity in this material on account of similarities with
spin-fluctuation induced superfluidity in $He^{3}$
\cite{Fay+Appel}.

Several theoretical studies
of the spin fluctuations of $Pd$ have been carried
out. The early ones were based on a jellium model for the
electrons and examined the contribution that spin fluctuations
make to the specific heat and their influence on any
superconducting phase transition as well as
describing the cross-section that would be measured in an
inelastic neutron scattering experiment \cite{Doniach}. Later work
has attempted to remove the deficiencies of such a single-band
model and to describe the band structure effects more
realistically \cite{SW+Sav}. From his calculations of the
inelastic neutron scattering cross-section for a range of
scattering angles and incident energies, Doniach \cite{Doniach}
estimated the
circumstances under which the paramagnons of $Pd$ might be
observed. Very recently time of flight measurements on a neutron
spallation source have succeeded in detecting these
long-wavelength, long-lived spin fluctuations \cite{Hayden}.

Using the formalism described in the last section, we have
calculated the temperature susceptibility of $Pd$ at $T=100$K and
have then determined the dynamic susceptibility $\chi({\bf q},w)$.
Currently our calculations use descriptions of the
equilibrium states in terms of scalar-relativistic atomic sphere
approximation (ASA) effective one-electron scattering potentials
and charge densities. The formalism, however, is applicable to full
potentials and non-spherically symmetric charge densities. The
lattice spacing of 3.83 $\dot{A}$
which we used in the $Pd$ calculations was
determined from the minimum of the
ASA total energy and found to be
1.5$\%$ smaller than the experimentally measured
value. At 15.7 $\mu_{B}^{2} eV^{-1}$ our calculated uniform
susceptibility $\chi (0,0)$ is somewhat lower than the value 24.3
$\mu_{B}^{2} eV^{-1}$ measured experimentally \cite{Hayden}.  For
small wavevectors ${\bf q}$ the inverse of our calculated static
susceptibility depends on the magnitude squared of ${\bf q}$, i.e.
$\chi^{-1}({\bf q},0) \approx \chi^{-1}(0,0) + c q^{2}$  as also
found experimentally and in other calculations
\cite{SW+Sav}. We obtain a value of approximately 900 $\mu_{B}^{2}
\dot{A}^{2} meV$ for $c$ whereas the value extracted from
inelastic neutron scattering experiments
is 294 $\pm$ 130 $\mu_{B}^{2} \dot{A}^{2} meV$. Since
palladium is so close to a ferromagnetic phase transition both
$\chi (0,0)$ and $c$ are very sensitive to the value of the lattice
parameter used and the RPA approximation used for the
exchange-correlation interactions.
Using the experimental lattice spacing in the calculations greatly
increases $\chi (0,0)$ (to 23.5 $\mu_{B}^{2} eV^{-1}$ close to the
value determined from experiment) and
both $\chi (0,0)$ and $c$ are likely to
be altered if spin fluctuation interaction effects are
treated beyond the LDA \cite{MD+Moriya+LT}.
Key aspects of the dynamic behaviour turn out not to
be so sensitive.

In figure 1 we show the calculated
imaginary part of $\chi({\bf q},w)$ for a range of frequencies
up to 200 meV for selected wavevectors ${\bf q}$ along the
$(1,0,0)$ direction. The sharp peaks at low frequencies for
$ q < 0.1$ indicate paramagnon-like behaviour. This is illustrated
further in figure 2 which shows the magnetic correlations $<
m^{2}({\bf q})>$ using energy cutoffs of 500meV and 50meV. Once
again the paramagnons are clearly visible for a narrow region of
small wave-vectors ${\bf q}$.

We find that the results
can be interpreted in terms of two overdamped simple harmonic
oscillators each with a characteristic timescale. The longer
timescale one encapsulates the paramagnon behaviour. $Im \chi ({\bf
q}, {\bf q},w)$ can now be written in the following way
\begin{equation}
Im \chi ({\bf q}, {\bf q},w) \approx \frac{\chi_{1}({\bf
q}) (w/\Gamma_{1}({\bf q}))}{(1+(w/\Gamma_{1}({\bf
q}))^{2})} +
\frac{\chi_{2}({\bf q}) (w/\Gamma_{2}({\bf
q}))}{(1+(w/\Gamma_{2}({\bf q}))^{2})}
\label{eq:2scale}
\end{equation}
Figure 3(a) shows the quantities $\chi_{1}({\bf
q})$, $\chi_{2}({\bf q})$ and their sum whereas figure 3(b) shows
$\Gamma^{-1}_{1}({\bf q})$ and
$\Gamma^{-1}_{2}({\bf q})$. Evidently at small $q$ the slow
relaxation process is dominant and
$\hbar/\Gamma_{1}({\bf
q})$ describes the long time taken for a spin density perturbation
to relax back to the equilibrium value. Moreover, we find that
the product of $\chi_{1}({\bf q})$ with
$\Gamma_{1}({\bf q})$ varies
as $\gamma_{0} q$ for small $q$, so that spin density disturbance
decays over a time proportional to the wavelength, i.e. exhibits
{\it Landau damping} \cite{GGLetc}. We find $\gamma_{0}$ to
be 2.13 $\mu_{B}^{2} \dot{A}$ which is in good agreement with the
experimentally determined value of 1.74 $\pm$ 0.8 \cite{Hayden} and
evidently fairly insensitive to the treatment of
exchange-correlation effects.

\section{Incommensurate antiferromagnetic spin
fluctuations in $Cr_{95}V_{5}$.}

Chromium, an early 3d transition metal, loses its
anti-ferromagnetic ground state when electrons are removed as
a suitable dopant is added. For example a
strongly exchange-enhanced paramagnet exhibiting
anti-ferromagnetic paramagnons is formed when just a few atomic percent
of chromium are substituted by vanadium. In this section we
explore the similarities and differences between these excitations
in one such system, $Cr_{95}V_{5}$,
and those reported for palladium. We also examine
the temperature dependence.

There is an extensive literature on chromium as the archetypal
itinerant anti-ferromagnet (AF) and its alloys \cite{Fawcett}.
It is well known that $Cr$'s
famous incommensurate spin density wave (SDW)
    ground state is linked to the nesting
    wave-vectors ${\bf q}_{nest}$ identified in the Fermi
surface. This feature also carries over to its
dilute alloys which have a range of AF properties \cite{Fawcett} (see
figure 4). Indeed the paramagnetic states of some
of these alloys have recently acquired a new relevance on
account of analogies drawn with the high temperature
superconducting cuprates especially
$(La_{c}Sr_{1-c})_{2}CuO_{4}$ \cite{Mason1992}. Starting with
`parent' materials $Cr_{95}Mn_{5}$ or $Cr_{95}Re_{5}$ which are
simple commensurate AF materials and then lowering the electron
concentration by suitable doping causes the Ne$\acute{e}$l
temperature, $T_{N}$,
to drop rapidly to zero. The paramagnetic metal
which forms for dopant concentrations
slightly in excess of the critical concentration for
this quantum phase transition is characterised
by incommensurate paramagnetic spin fluctuations. In
\cite{dyn} we described the spin fluctuations in the
paramagnetic state of three
representative systems
$Cr_{95}Re_{5}$, $Cr$ and $Cr_{95}V_{5}$. We tracked the
tendency for the spin fluctuations to become more nearly
commensurate with increasing frequency and also with increasing
temperature \cite{dyn} as observed in experimental data. Our
estimates of $T_{N}$ in $Cr_{95}Re_{5}$ and $Cr$ of 410K and
280K, respectively, were also in fair agreement with the experimental
values of 570K and 311K.

Although there have been several simple parameterised models
    to describe the magnetic properties of $Cr$ and its alloys
    \cite{model}, these have all
    concentrated on the approximately nested electron `jack' and
slightly larger octahedral hole pieces of the Fermi surface (FS)
\cite{Fawcett}.
These sheets can be seen clearly in
figure 4 which shows the Fermi surface of $Cr_{95}V_{5}$. The FS of
$Cr_{95}Re_{5}$'s is close to being perfectly nested and
we calculated in \cite{dyn}
its spin fluctuations above $T_{N}$ to be nearly commensurate for
small frequencies $w$. We also looked at $Cr$
and $Cr_{95}V_{5}$. $Cr$'s and $Cr_{95}V_{5}$'s Fermi
surfaces are progressively worse nested than $Cr_{95}Re_{5}$'s.
In our calculations for
these two systems \cite{dyn} we found the dominant slow
spin fluctuations to be incommensurate
with wave-vectors equal to
the FS nesting vectors ${\bf q}_{nest}$. $Cr_{95}V_{5}$'s is
shown on figure 4. At best, simple parameterised
models only include the effects
    of all the remaining electrons via an electron reservoir.
Whilst finding the obvious
    similarities from the FS basis
    between our results and results from such models, we showed that a
complete picture
    is obtained only when an electronic band-filling effect which
favours a
    simple AF
    ordering at low temperature is also considered.

As for the case of $Pd$ we found that
the spin
    fluctuations in the paramagnetic state of the three systems are
given an accurate description in
terms of a overdamped
diffusive simple
    harmonic oscillator model. Here, however, a single channel is
sufficient i.e. the susceptibility closely fits the following
\begin{equation}
Im \chi ({\bf q}, {\bf q},w) \approx \frac{\chi({\bf
q}) (w/\Gamma({\bf q}))}{(1+(w/\Gamma({\bf q}))^{2})}
\end{equation}

In this section we concentrate on the exchange-enhanced paramagnet
$Cr_{95}V_{5}$ and focus in particular on
the temperature dependence of the
incommensurate spin fluctuations and show how the
parameters of the oscillator model scale with temperature.

Figure 1(a) of reference \cite{dyn} showed $Im \chi ({\bf
q}, {\bf q},w)$ for wavevectors ${\bf q}$ along the $\{
0,0,1\}$ direction for frequencies up to 500meV. It showed,
in agreement with experiment \cite{Hayden}, incommensurate
AF paramagnons persisting up to high frequencies with
intensity comparable to that at low $w$. This is in
striking contrast to the paramagnons of $Pd$. Figure 5 shows
this persistence of the spin fluctuations in $Cr_{95}V_{5}$ to high
frequencies in a concise way. It shows
the magnetic correlations $<m^{2}({\bf q})>$ at 100K for
the same two energy cutoffs 50 and 500meV as was used to obtain
figure 2 for palladium. Unlike those
in the nearly ferromagnetic metal, the magnetic correlations
grow by more than an order of magnitude for all the wave-vectors
shown as the energy cutoff is extended by a similar factor.
The variation with temperature of the dynamic
susceptibility and the magnetic correlations have also been
investigated. Figure 6 shows $<m^{2}({\bf q})>$ calculated with a
energy cutoff of 500meV for 50K, 300K and 600K. There is a
general trend for the weight to transfer towards the commensurate
wavevector $(0,0,1)$ with increasing temperature. For a smaller
energy cutoff $<m^{2}({\bf q})>$, integrated over all
${\bf q}$, decreases with increasing temperature.

 Figures 7(a) and (b) show that both the static susceptibility
$\chi({\bf q})$ and the spin relaxation time $\Gamma^{-1}({\bf
q})/\hbar$
also peak at the incommensurate nesting vector $(0,0,0.9)$ at low
temperature and that weight is shifted towards $(0,0,1)$ at higher
temperatures.
Figure 7(b) when compared with the analogous figure 3(b)
for $Pd$ reveals the main difference between the dominant spin
fluctuations in the nearly antiferromagnetic $Cr_{95}V_{5}$ and the
nearly ferromagnetic $Pd$. For the dominant spin fluctuations, the
spin relaxation times are some
50 times faster. The small ${\bf Q}$ behavior is
different also. (${\bf
Q}={\bf q}-{\bf q}_{0}$, where ${\bf q}_{0}=0$ for $Pd$ and an
incommensurate wave-vector for $Cr_{95}V_{5}$.)
The product $\gamma ({\bf q})$ of $\chi ({\bf q})$ and
$\Gamma ({\bf q})$ tends to a constant for values of ${\bf q}$
close to ${\bf q}_{0}$ for $Cr_{95}V_{5}$ (consistent with
ideas that the dynamical critical exponent \cite{GGLetc} should be
2 for antiferromagnetic itinerant electron systems) unlike the
linear variation of $\gamma ({\bf q})$ with
$|{\bf q}|$ for small ${\bf q}$ found in $Pd$. Between 50K and
300K, $\gamma ({\bf q})$ shows a weak variation with temperature
where it has values of roughly 4 $\mu^{2}_{B}$ and 6.5
$\mu^{2}_{B}$ at the incommensurate wave-vector ${\bf q}_{0}$ and
${\bf q}=(0,0,1)$ respectively. Above 300K, $\gamma ({\bf q})$
loses the trough around ${\bf q}_{0}$ and by 600K it is roughly
constant at 5.5 $\mu^{2}_{B}$ for the range of ${\bf q}$ from
${\bf q}_{0}$ to $(0,0,1)$.

\section{Conclusions}

We have described the framework and method of implementation of
our scheme for carrying out ab-initio
calculations of the dynamic paramagnetic spin susceptibilities
of solids at finite temperatures. The approach is based upon
Time-Dependent Spin Density Functional Theory and is applicable to
compositionally disordered alloys with multi-atom per unit cell
crystal structures. (We note here that this approach may
also be adapted
to the study of magnetic excitations in magnetically ordered
materials.) From an imaginary time
multiple scattering Green function formalism an expression for the
temperature susceptibility has been derived and the techniques
appropriate for its evaluation described. We have shown how
the dynamic susceptibility is obtained from this by analytic
continuation from Matsubara frequencies in the complex plane
to the real frequency axis \cite{Godby}. This step provides a
natural interpretation of the spin dynamics
in terms of overdamped oscillator models.

Although ultimately aiming to investigate the spin fluctuations in
systems with complex lattice structures such as $Sr_{2}RuO_{4}$ and
$MnSi$ which are of topical interest, here we have compared and
contrasted the nearly ferromagnetic transition metal $Pd$ with the
nearly anti-ferromagnetic dilute $Cr$ alloy, $Cr_{95}V_{5}$. In
both cases we have been able to identify `slow' spin
fluctuations so that in due course
the mode-mode coupling amongst them may be incorporated into
spin-fluctuational theories which describe the low
temperature properties of these materials \cite{GGLetc,FSM-SF}.
For the case of $Pd$ the spin susceptibility can be
interpreted in terms of two oscillator models. One describes
spin fluctuations with fairly `fast' relaxations times but the
other describes the spin fluctuations for all wave-vectors
which have spin relaxation times
more than 40 times as slow as the time taken for an
itinerant d-electron to hop between lattice sites. Clearly
there is a natural time scale separation between these
spin fluctuational modes and the electronic
degrees of freedom. Consequently spin fluctuational theory
can be used with these modes without
the need for any wave-vector cutoff. Work is in
progress to carry this out. On the
other hand $Cr_{95}V_{5}$'s dynamic spin
susceptibility is interpreted
in terms of a single oscillator. Here the spin relaxation times are
only some 10 times slower than typical d-electron hopping times for
modes with ${\bf q}$'s in a limited region of
the Brillouin zone and
therefore mode-mode coupling effects may not be so important.
This may explain why our calculations \cite{dyn} of the static
susceptibility of this
$Cr$ alloy (and $Cr$ and its other dilute alloys) receive an
adequate description in terms of what is essentially an ab-initio
Stoner theory.

\section{Acknowledgements.}

We are grateful to S.Hayden and R.Doubble for useful discussions.
JBS and BG also acknowledge support from the Psi-k
Network funded by the ESF Programme
on `Electronic Structure Calculations for Elucidating the
Complex Atomistic Behaviour of Solids and Surfaces. JP acknowledges
support from the Thailand Research Fund through contract RTA/02/2542
and DDJ acknowledges support from the U.S. Department of Energy grants
DE-FG02-96ER45439 with the Fredrick Seitz Materials Research Laboratory
and DE-AC04-94AL85000 with Sandia.

\begin{figure}
\caption{$Im \chi({\bf q},{\bf q},w)$ of $Pd$ at $T=$100K in units
of $\mu_{B}^{2} eV^{-1}$ for wave-vectors ${\bf q}$ along the
$\{ 0,0,1 \}$ direction where ${\bf q}$ is in units of $\pi/a$
(a is the lattice spacing). The frequency axis, $w$, is marked
in meV. The full curve is for ${\bf q}=$(0,0,0.02), then for curves
with decreasing dash-length, ${\bf q}=$ (0,0,0.1),(0,0,0.2) and
(0,0,0.5). }
\end{figure}
\begin{figure}
\caption{The variance of the spin fluctuations in $Pd$ at $T=$100K,
$<m^{2}({\bf q})>$, in $\mu_{B}^{2}$ for
  wave-vectors ${\bf q}$ along the $\{ 0,0,1 \}$ direction with
energy cutoffs of 500 meV (full line) and 50 meV (dashed line).}
\end{figure}
\begin{figure}
\caption{(a) The static susceptibility of $Pd$ at
$T=$100K partitioned into 2 parts
$\chi_{1}({\bf q})$, $\chi_{2}({\bf q})$ of equation
~(\ref{eq:2scale}) in units of
$\mu_{B}^{2} eV^{-1}$ for wave-vectors ${\bf q}$ along the $\{
0,0,1 \}$ direction. $\chi_{1}({\bf q}) + \chi_{2}({\bf q})$ is
shown by the full line, $\chi_{1}({\bf q})$ by the  long-dashed
line and $\chi_{2}({\bf q})$ by the short-dashed line. (b) The time
scales from eq.~(\ref{eq:2scale})
(divided by $\hbar$) in $eV^{-1}$,
$\Gamma^{-1}_{1}({\bf q})$ (full line), $\Gamma^{-1}_{2}({\bf q})$ (dashed line)
for $Pd$ at
$T=$100K. {\it Note $\Gamma^{-1}_{2}({\bf q})$ has been scaled up by 100
for display purposes.}}
\end{figure}
\begin{figure}
\caption{The Fermi surface
(FS) of $Cr_{95}V_{5}$ from
Bloch spectral function calculations in the $q_{z}=0$ plane. The
dashed arrows mark the nesting vector ${\bf q}_{nest}$
connecting the $H$-centered hole octahedron to the slightly smaller
$\Gamma$-centered electron surface.
The FS is well defined with only a small
disorder broadening $< 0.01\pi/a$.}
\end{figure}
\begin{figure}
\caption{The variance of the spin fluctuations in
$Cr_{95}V_{5}$ at $T=$100K, $<m^{2}({\bf q})>$, in $\mu_{B}^{2}$ for
wave-vectors ${\bf q}$ along the $\{ 0,0,1 \}$ direction with
energy cutoffs of 500 meV (full line) and 50 meV (dashed line).}
\end{figure}
\begin{figure}
\caption{The variance of the spin fluctuations in
$Cr_{95}V_{5}$,
$<m^{2}({\bf q})>$, at $T=$50K (full line),
300K (long-dashed) and 600K (short-dashed), in $\mu_{B}^{2}$ for
wave-vectors ${\bf q}$ along the $\{ 0,0,1 \}$ direction with a
energy cutoff of 500 meV.}
\end{figure}
\begin{figure}
  \caption{(a) The static susceptibility of $Cr_{95}V_{5}$ at
$T=$50K (full line),
  300K (long-dashed) and 600K (short-dashed), in $\mu_{B}^{2}
eV^{-1}$ along the $\{ 0,0,1 \}$ direction. (b) The time scale
(divided by $\hbar$) in $eV^{-1}$, $\Gamma^{-1}({\bf q})$ for
$Cr_{95}V_{5}$ at the same temperatures.}
\end{figure}
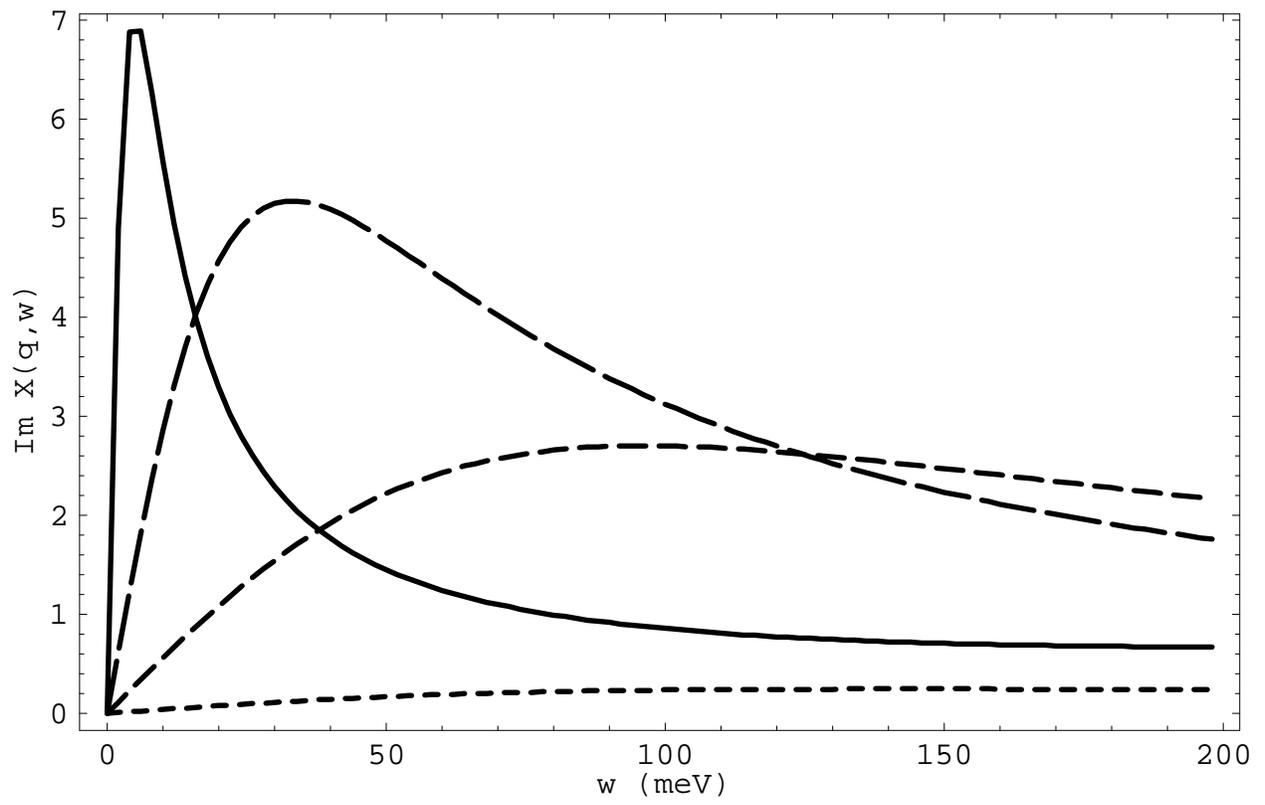
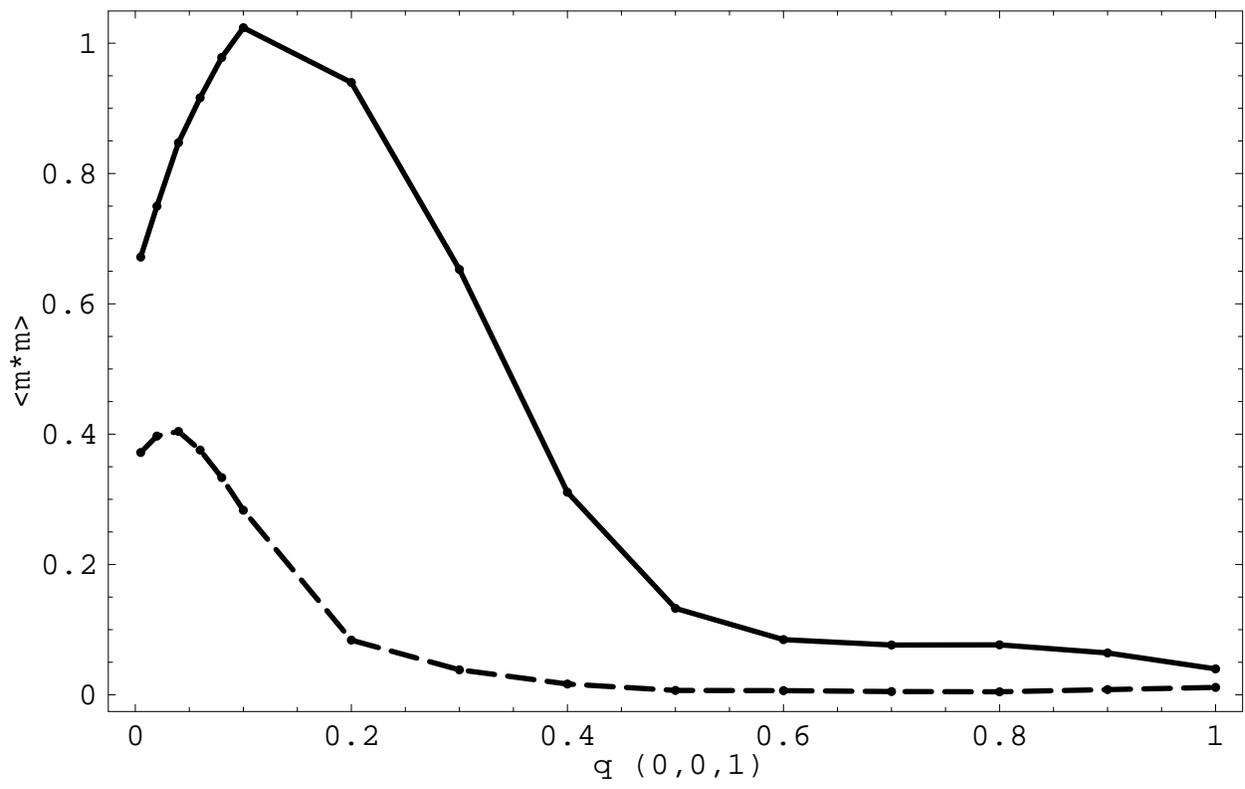
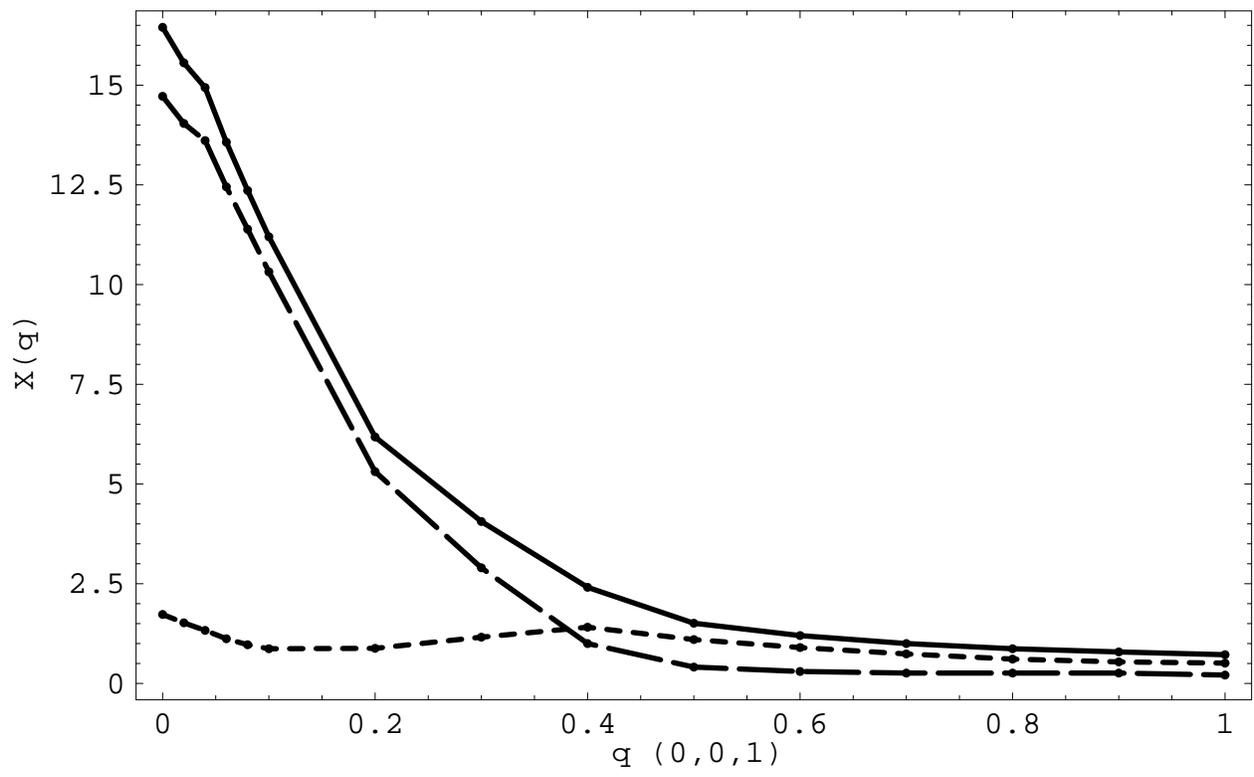
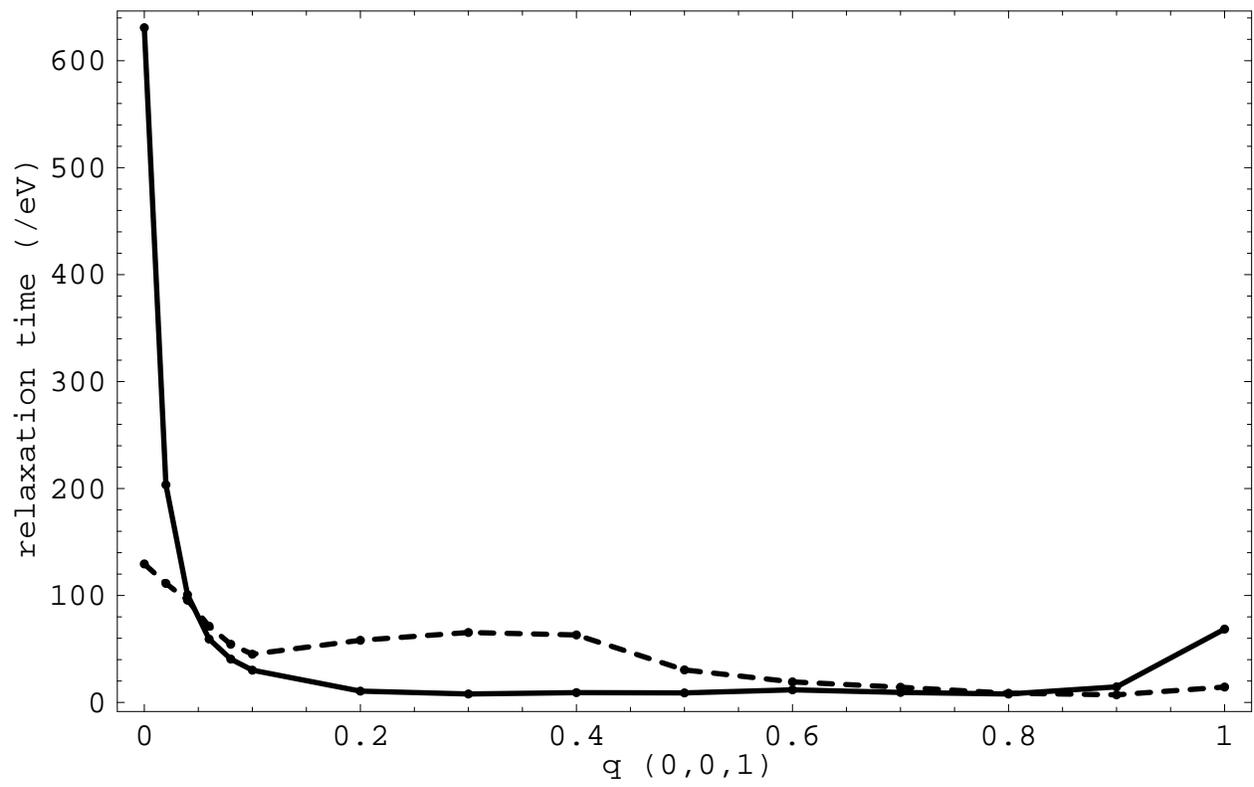
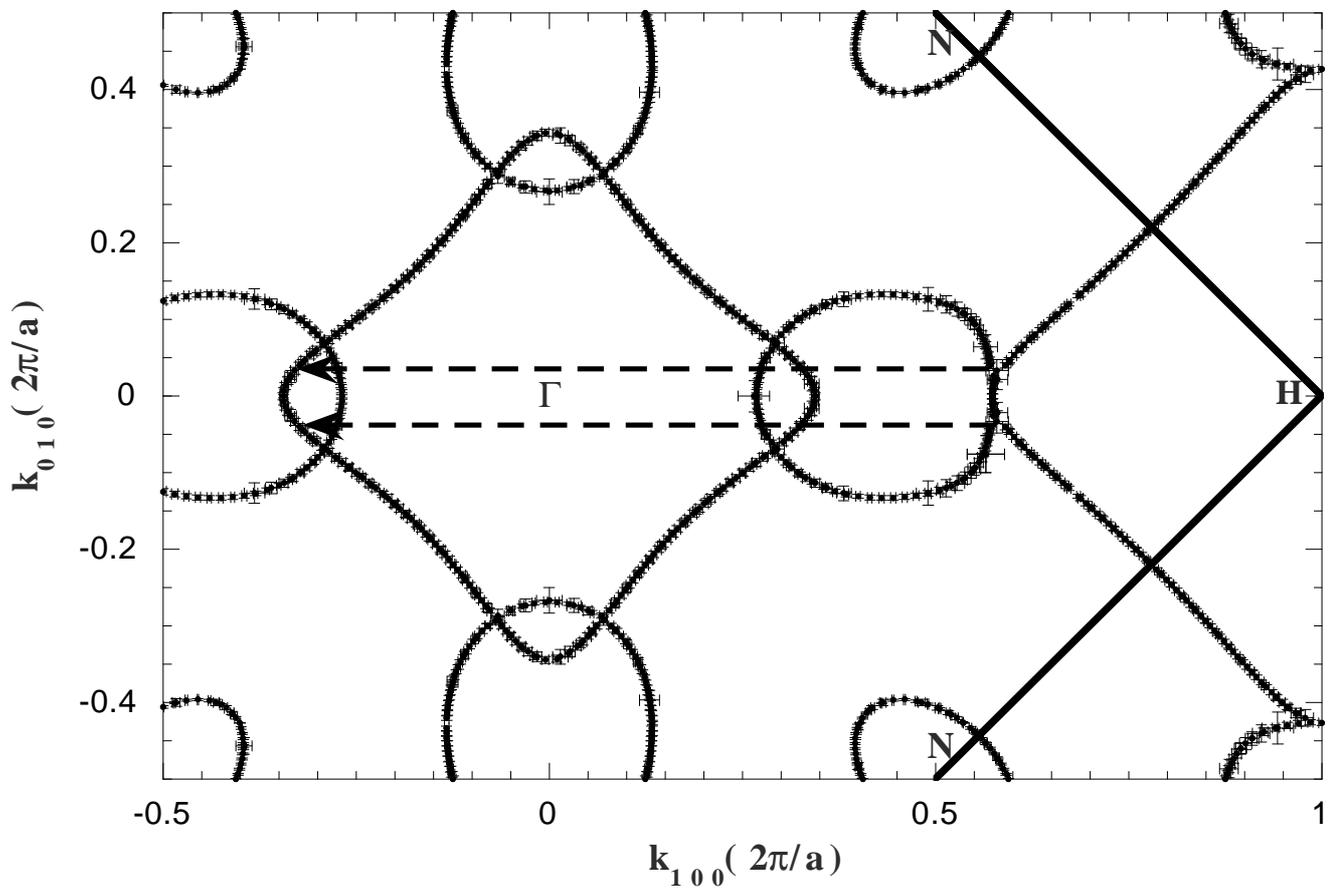
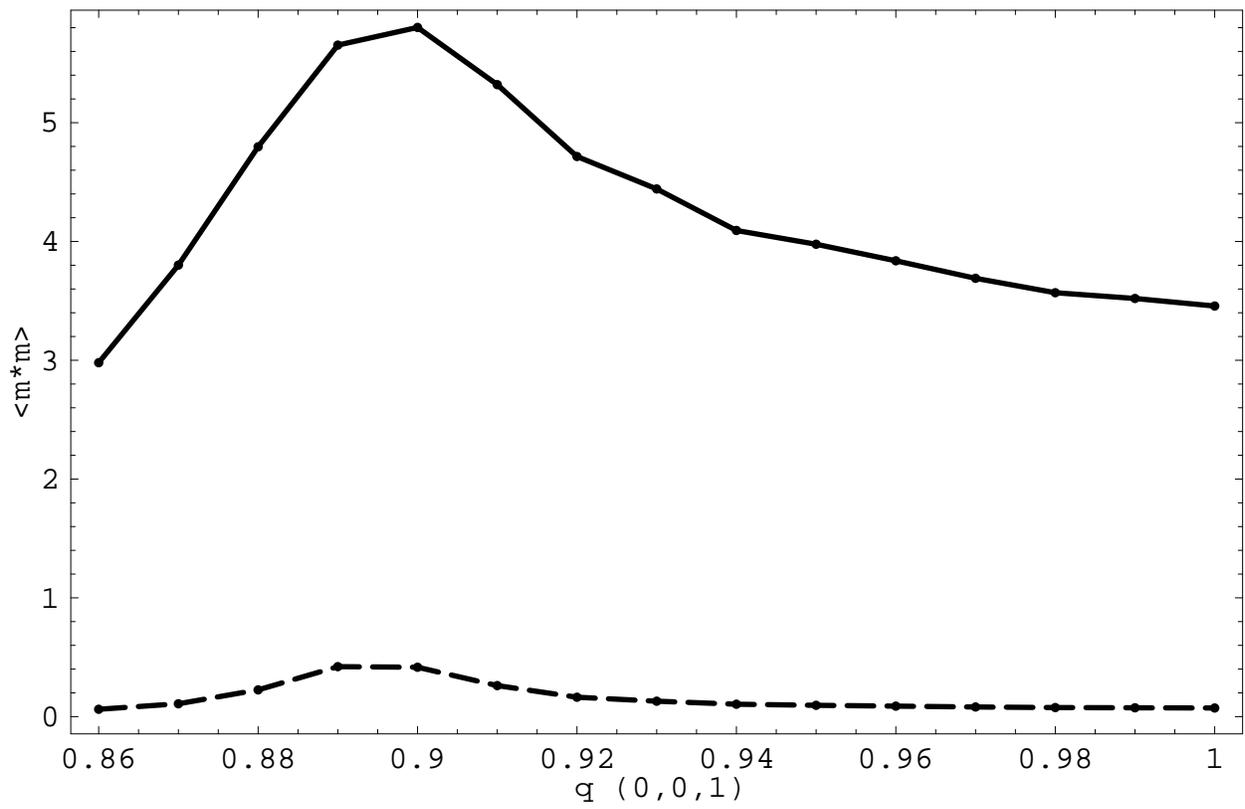
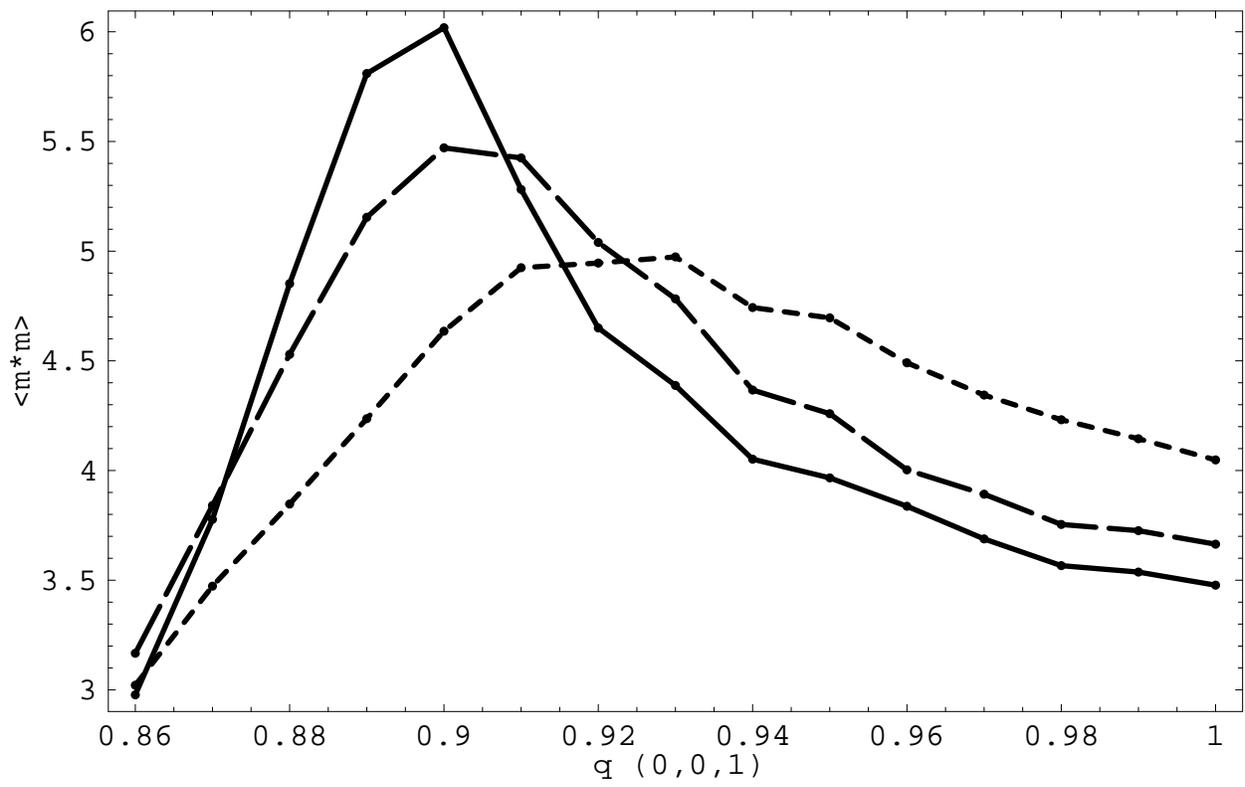
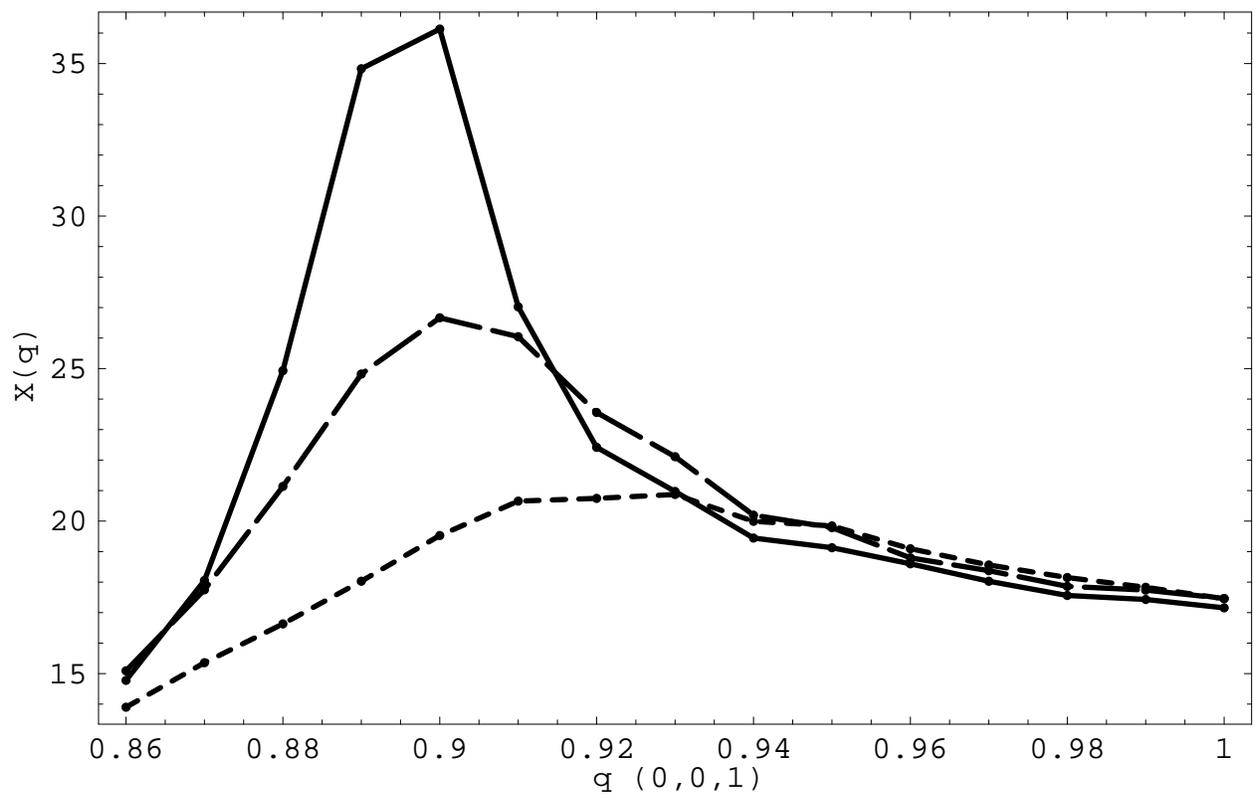
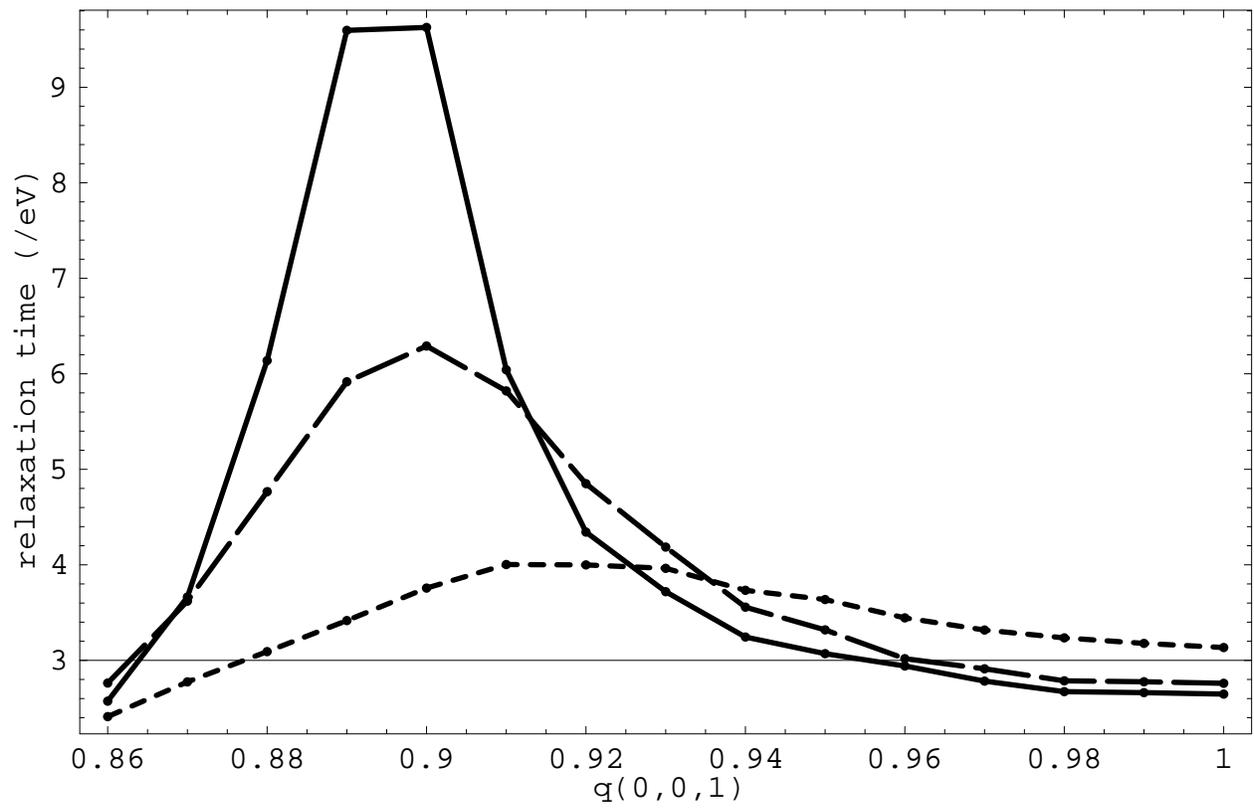
\end{document}